\documentclass[fleqn,usenatbib]{mnras}

\usepackage[T1]{fontenc}
\usepackage{ae,aecompl}

\usepackage{graphicx}	
\usepackage{amsmath}	
\usepackage{amssymb}	
\usepackage{subfig}
\usepackage{tablefootnote}

\newcommand\T{\rule{0pt}{2.6ex}}
\newcommand\B{\rule[-1.6ex]{0pt}{0pt}}

\title[GAMA: The $325$ MHz Radio Luminosity Function]{Galaxy And Mass
  Assembly (GAMA): The $325$ MHz Radio Luminosity Function of AGN and Star
Forming Galaxies}

\author[Matthew~Prescott et al.]
{Matthew~Prescott$^{1}$\thanks{E-mail:~mxp@uwcastro.org},
T.~Mauch$^{2}$,
M.J.~Jarvis$^{1,3}$, 
K.~McAlpine$^{1}$,
D.J.B.~Smith$^{4}$,
\newauthor
S.~Fine$^{1}$,
R.~Johnston$^{1}$,
M.J.~Hardcastle$^{4}$,
I.K.~Baldry$^{5}$,
S.~Brough$^{6}$,
\newauthor
M.J.I.~Brown$^{7}$,
M.N.~Bremer$^{8}$,
S.P.~Driver$^{9,10}$,
A.M~Hopkins$^{6}$,
L.S.~Kelvin$^{5}$,
\newauthor
J.~Loveday$^{11}$,
P.~Norberg$^{12}$,
D.~Obreschkow$^{9}$ and 
E.M.~Sadler$^{13}$
\\
$^{1}$Department of Physics and Astronomy, University of the Western Cape, Private Bag X17, Bellville 7535, South Africa
\\
$^{2}$SKA Africa, Third Floor, The Park, Park Road, Pinelands 7405, South Africa
\\
$^{3}$Oxford Astrophysics, Denys Wilkinson Building, Keble Road,
Oxford OX1 3RH, UK
\\
$^{4}$Centre for Astrophysics Research, Science \& Technology Research
Institute, University of Hertfordshire, Hatfield, Herts, AL10 9AB, UK
\\
$^{5}$ Astrophysics Research Institute, Liverpool John Moores University, IC2, Liverpool Science Park, 146 Brownlow Hill, Liverpool, L3 5RF
\\
$^{6}$ Australian Astronomical Observatory, PO Box 915, North Ryde,
NSW 1670, Australia
\\
$^{7}$ School of Physics and Astronomy, Monash University, Clayton, Victoria 3800, Australia
\\
$^{8}$ H.H. Wills Physics Laboratory, University of Bristol, Tyndall Avenue, Bristol BS8 1TL, UK
\\
$^{9}$ ICRAR, University of Western Australia, 35 Stirling Highway, Crawley, WA 6009, Australia
\\
$^{10}$ SUPA, School of Physics and Astronomy, University of St Andrews, North Haugh, St Andrews, KY16 9SS, UK
\\
$^{11}$ Astronomy Centre, University of Sussex, Falmer, Brighton BN1
9QH, UK
\\ 
$^{12}$ Institute for Computational Cosmology, Department of Physics,
Durham University, South Road, Durham DH1 3LE, UK
\\
$^{13}$ School of Physics, University of Sydney, NSW 2006, Australia
}

\date{Accepted 2015 December 30. Received 2015 December 24; in
  original form 2015 August 27.}

\pubyear{2015}

\begin{document}
\label{firstpage}
\pagerange{\pageref{firstpage}--\pageref{lastpage}}
\maketitle

\begin{abstract}
Measurement of the evolution of both active galactic nuclei (AGN) and
star-formation in galaxies underpins our understanding of galaxy evolution
over cosmic time. Radio continuum observations can provide key
information on these two processes, in particular via the mechanical
feedback produced by radio jets in AGN, and via an unbiased dust-independent
measurement of star-formation rates.
  In this paper we determine radio luminosity functions at 325 MHz for a sample of 
  AGN and star-forming galaxies by matching a 138 deg$^{2}$ radio survey conducted
  with the Giant Metrewave Radio Telescope (GMRT), with optical
  imaging and redshifts from the Galaxy And Mass Assembly (GAMA) survey. 
  We find that the radio luminosity function at 325 MHz for star-forming galaxies 
  closely follows that measured at 1.4 GHz. 
  By fitting the AGN radio luminosity function out to $z=0.5$ as a double power law, and
  parametrizing the evolution as $\Phi \propto (1+z)^{k}$,  we find
  evolution parameters of $k = 0.92 \pm 0.95$ assuming pure density evolution and $ k = 2.13 \pm
 1.96$ assuming pure luminosity evolution. 
 We find that the Low Excitation Radio Galaxies are the dominant population in space density at lower
  luminosities.     
  Comparing our 325~MHz observations with radio continuum imaging at
  1.4~GHz, we determine separate radio luminosity functions for steep and
  flat-spectrum AGN, and show that the beamed population of flat-spectrum sources in
  our sample can be shifted in number density and luminosity to
  coincide with the un-beamed population of steep-spectrum sources, as is expected in the orientation based unification 
  of AGN.       
\end{abstract}

\begin{keywords}
surveys -- galaxies:luminosity function -- galaxies: evolution -- galaxies:
formation -- radio continuum: galaxies -- galaxies: active 
\end{keywords}

\section{Introduction}
\label{intro}
   
In recent years it has become apparent that Active Galactic Nuclei
(AGN) play an important role in the formation and evolution of
galaxies. Every massive galaxy is now thought to contain a central
supermassive black hole that can undergo periods of rapid gas
accretion to produce an AGN. The accretion activity of AGN may also
have a strong
interplay with star formation. Indirect evidence for this comes from comparisons made
between the evolution of AGN activity and the star-formation rate of
the Universe, which have been shown to follow each other over cosmic history
\citep[e.g.][]{Franceshini1999}. In the local Universe more evidence comes
from the correlation between the black hole and bulge masses of
galaxies \citep{Kormendy1995, Magorrian1998}. 
Although more recent results have shown this relationship deviates
from being linear when probing lower masses \citep{Graham2015}, the
relationship for more massive galaxies is thought to arise from
the regulation of star formation in the bulge due
to `AGN feedback' \citep [for a review see][]{Fabian2012}. AGN feedback has also been incorporated into
semi-analytical models \citep{Croton2006, Bower2006}, as a way to
quench star formation and produce the observed colour bimodality of
galaxies \citep{Strateva2001, Baldry2004}.  
 
The radio source population probed by the current generation of radio surveys probes a mixture of
galaxies powered by star-formation and AGN activity. At 1.4~GHz flux densities of $\sim 10$ mJy and
above, the majority of radio sources are radio-loud AGN. Below $10$~mJy 
there is an increasing number of star-forming galaxies \citep{Condon1989}. \citet{Fanaroff1974} found that radio-loud
AGN could be subdivided into two types (known as Fanaroff-Riley
(FR) types I and II), based on their radio morphologies, which were
found to show a division at approximately $L_{\rm{178-MHz}} =10^{25}$
W~Hz$^{-1}$. Those having brighter cores and diffuse lobes are labelled as 
FR I sources and those with highly collimated jets,
which produce `hot-spots' of high surface brightness at their edges, as FR IIs.    

A different classification system based on the presence or absence of
narrow emission lines in the optical spectra of radio-loud AGN has also been used over
the years \citep {Hine1979, Laing1994, Jackson1997,
  Willott2001}. Those with high-excitation emission lines
being referred to as High Excitation Radio Galaxies (HERGs) and those
without as Low Excitation Radio Galaxies (LERGs). HERGs are the
dominant population above $L_{\rm{1.4-GHz}}=10^{26}$ W~Hz$^{-1}$ \citep{Best2012} and are more likely
to be associated with the more powerful FR~II sources, compared to the
LERGs which are more associated with FR~Is (although not exclusively).
The dichotomy in these populations is believed to be due to two
different modes of accretion, with the LERG population undergoing
radiatively inefficient accretion of hot gas from the galaxies'
interstellar medium and surrounding galaxies, and the HERG
population undergoing radiatively efficient accretion of cold gas by
mergers and interactions between the host galaxy and gas rich
systems \citep{Allen2006, Evans2006, Hardcastle2007, Smol2009, Heckman2014, Fernandes2015}. Both accretion
modes are thought to be important processes in the AGN feedback
mechanism. In order to fully understand these different mechanisms,
the evolution of these populations has to be determined, usually
conducted by either a
$V/V_{\rm max}$ analysis or by producing radio luminosity functions,
whose evolution is parametrized in some way. 

It is now established that the most powerful
radio-loud AGN (with $L_{\rm{1.4-GHz}} > 10^{25}$ W~Hz$^{-1}$)
evolve strongly over the course of cosmic history, with studies
finding an increase in their number densities by a factor of 1000 from
$z \sim 0 - 2$ \citep{Longair1966,Laing1983, Dunlop1990, Willott2001}, which then
declines beyond $z \sim 3$ \citep{Shaver1996, Jarvis2000, Jarvis2001,Wall2005,Rigby2011}.
At lower luminosities, however, the picture is much more uncertain.

In an early study by \cite{Laing1983}, observations at 178 MHz
conducted as part of the 3CRR survey, revealed that sources with
$L_{\rm{178-MHz}}<10^{26}$ W~Hz$^{-1}$ exhibit no evolution via a $V/V_{\rm max}$ analysis.    
\cite{Jackson1999} also found little or no evolution in the FR I population,
compared to FR IIs at frequencies of 151 MHz. The trend that 
low luminosity sources ($L < 10^{25}$ W~Hz$^{-1}$) evolve much less than
higher luminosity sources was also seen by 
\cite{Waddington2001}. At 325 MHz, \cite{Clewley2004} found no significant
evolution for sources with $L_{\rm{325-MHz}} < 10^{26.1}$ W~Hz$^{-1}$~sr$^{-1}$ out to
$z=0.8$, from matching radio data from the Westerbork Northern Sky
Survey \citep[WENSS;][]{Rengelink1997} with optical SDSS
DR1 data and performing a $V/V_{\rm max}$ analysis.

These results are in contrast to those of \cite{Brown2001} who found strong
evolution in the low luminosity radio population at 1.4 GHz out to $z =0.55$, and
by assuming pure luminosity density evolution of the form $L \propto
(1+z)^{k}$ found $ 3 < k < 5$ for AGN with $10^{23} < L_{\rm{1.4-GHz}} < 10^{25}$ W~Hz$^{-1}$.
\cite{Sadler2007} also found significant evolution for AGN with
$10^{24} < L_{\rm{1.4-GHz}} < 10^{25}$ W~Hz$^{-1}$
consistent with pure luminosity evolution where $L \propto
(1+z)^{2.0 \pm 0.3}$ from $z = 0.7$, using the 2SLAQ \citep{Cannon2006} catalogue combined with 
Faint Images of the Radio Sky at Twenty-Centimeters
\citep[FIRST,][]{Becker1995} and NRAO VLA Sky Survey
\citep[NVSS,][]{Condon1998}.

More recent studies, probing fainter fluxes, have found milder evolution in
low-luminosity AGN. \cite{Smol2009} produced 1.4 GHz luminosity functions
using the VLA-COSMOS survey for AGN with $10^{21} < L_{\rm{1.4-GHz}} < 10^{26}$ W~Hz$^{-1}$ to $z =1.3$
and found modest evolution, with $L \propto (1 + z)^{0.8 \pm 0.1}$, or $\Phi \propto (1 + z)^{1.1 \pm 0.1}$,
assuming pure luminosity and density evolution respectively.

\cite{McAlpine2011} found that low-luminosity sources evolve
differently from their high-luminosity counterparts out to a redshift
of $ z = 0.8$ and the measured radio luminosity function was found to be consistent with an
increase in the comoving space density of low-luminosity sources by a
factor of 1.5.

\cite{Simpson2012} using deep $S_{\rm 1.4} > 100~\mu$Jy radio imaging in the Suburu/{\it XMM-Newton
Deep Field}, produced 1.4 GHz luminosity functions divided into the radio-loud and
radio-quiet AGN populations. They found that radio-quiet population
evolves more strongly than the radio-loud population. Again very little or no evolution in the
number density of the radio-loud AGN with $L_{\rm{1.4-GHz}} \le 10^{24} $ W~Hz$^{-1}$
out to $z = 1.5$ was observed.

Using 1.4 GHz VLA data combined with
photometric redshifts from the VIDEO survey, \cite{McAlpine2013} found that AGN evolve as 
$L\propto (1 + z)^{1.18 \pm 0.21}$ at relatively low radio luminosities,
assuming pure luminosity evolution. 

Finally, using deep from deep ($\sim 30 \mu $Jy) 1.4 GHz VLA observations of the {\it Chandra} Deep
  Field South (CDFS) \cite{Padovani2015} find that the number density of
  radio loud AGN evolves as $\propto (1+z)^{2}$ out to $z = 0.5$,
  after which it declines steeply as $\propto (1+z)^{-4}$.     

As well as tracing the population of AGN, radio emission is produced
from electrons in the H{\sc ii} regions near massive stars (free-free emission)
and cosmic ray electrons produced by supernova remnants (synchrotron emission), and offers
a probe of recent star-formation in galaxies without the need for an uncertain extinction correction
\citep{Condon1992}.    
It has become well established that the star-formation
rate of the Universe has declined by a factor of $\sim 10$, from $z=1.0$ to
$z=0.0$ \citep{Lilly1996, Madau1996, Hopkins2006, Madau2014}. Studies such as
\cite{Hopkins2004} for example, 
incorporated 1.4~GHz measurements into a compilation of star-formation
indicators to show that the star-formation rate in galaxies evolves
as $L \propto (1 + z)^{2.7 \pm 0.6}$ for pure luminosity evolution and $\phi \propto (1 + z)^{0.15 \pm 0.6}$ for density
evolution. \cite{Padovani2011} also found that star forming galaxies evolve as $L \propto (1+z)^{2.89 \pm 0.1}$, out to $z =2.3$, at
$1.4$ GHz using VLA observations of the {\it Chandra} Deep Field South
(CDFS). More recently, \cite{McAlpine2013} measured the 1.4 GHz radio luminosity
function of star-forming galaxies, from VLA data combined with
photometric redshifts from the VIDEO survey. They measured evolution in the star-formation rate density 
as $L \propto (1+ z)^{2.47 \pm 0.12}$ out to $z =1$.           

In this paper we present a measurement of the evolution of
radio-loud AGN and star forming galaxies to $z = 0.5$, by matching a catalogue of
radio sources measured at a frequency of 325 MHz from the Giant
Metrewave Radio Telescope (GMRT) to their optical counterparts in the
Galaxy And Mass Assembly (GAMA) survey. The 325~MHz GMRT survey is the deepest
available to date, over a sky area significant enough to alleviate the effects of cosmic
variance. The GAMA survey provides reliable spectroscopic redshifts, allowing
us to spectroscopically classify AGN and star-forming galaxies, in
addition to providing more accurate redshift and luminosity estimates than studies using
photometric redshifts. At present, studies of the evolution in the radio source population
made from samples selected at frequencies below 1.4~GHz have primarily relied on radio data
with a relatively shallow flux density limit and on photometric
redshifts in the optical \citep[e.g.][]{Clewley2004}.
Radio-loud AGN samples selected at lower radio frequencies are of interest
because the detected population is less dependent on the orientations of the jets.
The steep-spectrum lobes of radio galaxies dominate at lower radio frequencies, 
whereas in the GHz regime the doppler boosted flat-spectrum cores of
pole-on sources are more likely to be detected \citep{Jarvis2002}.

The structure of this paper is as follows. In Section~2 we outline the
GAMA survey and the GMRT data used in this study. In Section~3 we
explain the technique to match the radio sources to their optical
counterparts. In Section~4 we describe the way we have classified AGN
and star forming galaxies. In Section~5 we highlight some of the sample
properties and show how the spectral indices of the sources vary as a function of luminosity and redshift.
 In Section~6 we present our radio luminosity functions, and determine
 the evolution in low-luminosity radio sources. We go on to produce
 RLFs for the HERG and LERG populations, and RLFs for AGN
with steep and flat spectral indices. In Section~7 we compare our results to
other studies. Finally in Section~8 we summarise our main results.

Throughout this paper we assume $H_{0} = 70\,{\rm km s}^{-1}\,{\rm
  Mpc}^{-1}$, $\Omega_{m} = 0.3$ and $\Omega_{\Lambda} = 0.7$. 
                  
\section{Data}

\subsection{Galaxy And Mass Assembly} 

GAMA is a multi-wavelength (far-UV to radio) survey of $\sim 290\,000$
galaxies selected to be complete to $r = 19.8$ mag, combining photometry and spectroscopy from the latest wide-field
survey facilities \citep{Driver2009, Driver2011, Liske2015}. Covering $\sim 290$
deg$^{2}$ and probing galaxies to $z \sim 0.5$, GAMA allows the study of
galaxies and cosmology on scales between 1 kpc
 and 1 Mpc and provides
the link between wide-shallow surveys, such as the SDSS Main Galaxy
Sample \citep[SDSS MGS]{Strauss2002} and 2dFGRS \citep{Colless2001},
and narrow-deep surveys such as DEEP2 \citep{Davis2003}, $z$COSMOS
\citep{Lilly2007} and VVDS
\citep{LeFevre2005}.

Optical spectroscopy for the GAMA survey was conducted at the 3.9-m
Anglo-Australian Telescope (AAT) using the AAOmega spectrograph
\citep{Sharp2006} on 210 nights, over 6
years, between January 2008 and September 2014. For further details about spectroscopic
 target selection and the tiling strategy used for GAMA, the reader is
referred to \cite{Baldry2010} and \cite{Robotham2010} respectively. In
brief, galaxies are selected for spectroscopy using an input catalogue
drawn from the Sloan Digital Sky Survey (SDSS) Data Release 7
\citep{SDSSDR7} and UKIRT Infrared Deep Sky Survey (UKIDDS)
\citep{Lawrence2007}. Raw spectra are reduced and calibrated using a
pipeline described in \cite{Hopkins2013}. 
 GAMA makes extensive use of SDSS photometry, which is obtained for five broad-band filters ({\it ugriz})
 using a dedicated 2.5-m telescope at Apache Point, New Mexico,
 equipped with a mosaic CCD camera \citep{Gunn1998} and calibrated with a 0.5-m telescope
 \citep{Hogg2001}. For greater detail regarding the SDSS the reader is
 referred to \cite{York2000} and \cite{Stoughton2002}. 

In the following analysis we use imaging data for galaxies which make up the
{\it r}-band limited Main Survey, observed in three $12 \times 5$ deg$^{2}$
GAMA fields located along the celestial equator at 9h, 12h and 14.5h
(known as G09, G12 and G15). In order to measure the radio luminosity
function and its evolution, we select galaxies from GAMA TilingCat43 with redshifts
between $0.002 < z < 0.6$, $nQ \ge 3$, $r \le 19.8$ mag and a Survey Class $\ge 4$ which ensures
we are using reliable redshifts and have a well defined selection
limit in which we are $>98$ per cent complete. This results in a
sample of $185,125$ galaxies over G09, G12 and G15.     

\subsection{GMRT data}

The 325~MHz radio survey covering the GAMA fields, using the Giant
Meterwave Radio Telescope (GMRT), is fully described in
\cite{Mauch2013}\footnote{See also: http://www.extragalactic.info/mjh/gmrt/}. A total of 212, 15 minute pointings were observed over 8 nights in January 2009, 3 nights in May
2010 and 1 night in June 2010. These pointings overlap with 138 deg$^{2}$
of the GAMA G09, G12 and G15 regions. Flagging, calibration, self-calibration and
source detection of each pointing was conducted using a pipeline making
use of the AIPS software package. After reduction, the pointings in each field were
mosaiced, producing images with resolutions of $14''$, $15''$ and $23.5''$ with 
minimum rms noises of $\sim$ 0.8 mJy beam$^{-1}$, $\sim$ 1 mJy beam$^{-1}$
and $\sim$ 1.5 mJy beam$^{-1}$, for fields G09,
G12 and G15, respectively. A source catalogue was produced by fitting
elliptical Gaussians in the mosaics, which resulted in a final
catalogue that contains $5,264$ sources. 
To remove spurious detections, we limited the GMRT catalogue to include only sources brighter than
5$\sigma$ (Peak flux density/local RMS $\ge 5$), resulting in a catalogue of $4\,931$ radio
sources.  

\section{GAMA/GMRT Cross-Matching}

Cross-matching between the GAMA and the GMRT
catalogues was conducted by visual inspection of radio contours onto optical images.
This was done because the elliptical Gaussian model fitted to GMRT radio sources 
is insufficient to describe the wide range of source morphologies at the resolution of the
survey and the Gaussian fits are also sometimes affected by image artefacts. Automated cross-matching 
of the catalogues based on position separation will be unable to take account of these errors. 
We treat the cross-matching of single and multiple component radio sources
separately, where radio sources within a
radius of $50''$ of each other are considered as being potential multiple component
sources. Sources with no other radio source within $50''$ are treated as single component
sources. Using this method we are considering every potential
  optical counterpart from GAMA within $50''$ of a GMRT radio source.       

\subsection{Single Component Sources}

There are $3\,616$ single component radio sources (i.e. sources with no other radio source within $50''$) in the GMRT catalogue. 
In order to find optical counterparts, each radio source is matched to the nearest object in
the selected GAMA spectroscopic catalogue with a maximum offset of $15''$. 
Contour maps of each GMRT source were then produced and
overlaid on top of a $3'\times 3'$ SDSS $r$-band images of the
potential counterpart, and inspected to establish whether the radio source has a robust optical counterpart. 
Visual inspection was performed by two of us to ensure its reliability.
As a double-check to ensure 
the GMRT indentifications were real, contour maps from the 1.4 GHz surveys FIRST
and NVSS, where available, were also overlaid on the SDSS image. The FIRST and NVSS contours complement the GMRT data when matching, 
as FIRST provides resolved ($6.4''\times5.4''$ resolution in the GAMA regions) detections of the sources and
the NVSS, despite its lower resolution of $45''$, has greater surface brightness
sensitivity.
 
Of the $3\,616$ single component sources, we found $573$ lie within $15''$
of a GAMA source in the selected spectroscopic catalogue. 
We then visually matched $370$ out of these GMRT sources with their optical counterparts in GAMA.

For comparison, we also looked for single component counterparts using the likelihood ratio technique 
\citep{Sutherland1992, Smith2011, McAlpine2012}. This resulted in $330$ GAMA/GMRT matches with
reliabilities $> 0.8$. This method resulted in fewer matches overall, mainly due to the non-trivial structure of many radio
sources at the GMRT resolution \citep[e.g.][]{McAlpine2012}, so
we use our visually classified sample for the rest of this
paper. At present visual classifcation is acknowledged as being the most reliable
method of cross matching radio source catalogues with optical datasets \citep[e.g.][]{Fan2015}.

\subsection{Multiple Component Sources}

To find the counterparts of multiple component sources (i.e. radio sources with a radio counterpart within $50''$), 
GMRT contour maps of all radio sources were again overlaid with FIRST and NVSS contours, on top of a $5' \times
5'$ SDSS $r$-band image of the nearest GAMA object. Visual inspection of candidate multiple component radio sources
was performed in the same way as was done for single component sources.

From the initial GMRT catalogue of $4\,931$ $5\sigma$ radio source components, we
found $757$ potential multiple component sources.  
After inspection we identified $84$ multiple component radio sources,
together with a further $46$ objects that turned out to
be separate single component sources near unrelated GMRT
sources that could not be matched (bringing the total number of single component
sources to $414$).
The flux densities of multiple component objects are simply the sum
of the flux density of each individual component in the GMRT
catalogue. 

\begin{figure}
\includegraphics*[width=0.47\textwidth]{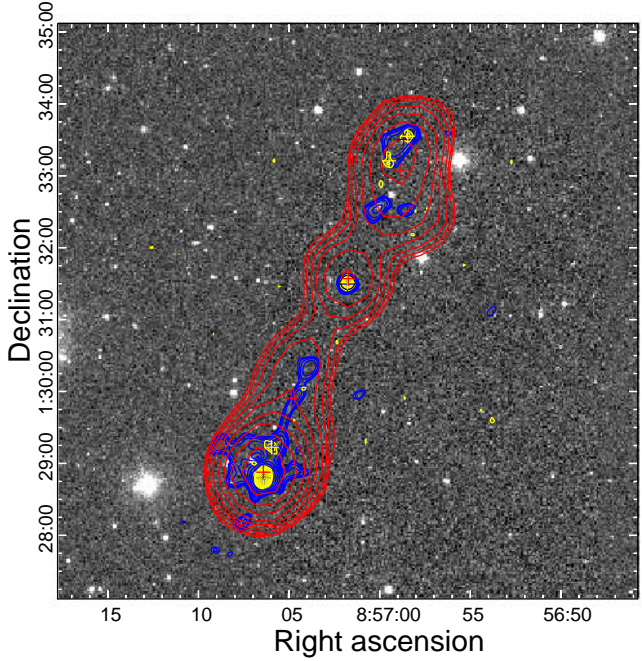}
\caption{An example of one of the GMRT/GAMA matches
  GAMAJ085701.76+013130.9, showing the SDSS $r$-band image, overlaid with radio contours from
  the GMRT (blue), FIRST (yellow) and NVSS (red). At z = 0.27, the
  FR II AGN has a projected size of 1.3 Mpc, making it a previously undiscovered giant radio
  galaxy (GRG).}
\label{GRG}
\end{figure}

Figure~\ref{GRG} shows an example of one of the GAMA/GMRT matches, 
GAMAJ085701.76+013130.9, with the SDSS
$r$-band image overlaid with radio contours from
the GMRT (blue), FIRST (yellow) and NVSS (red). At $z =0.27$ this
galaxy has a projected size of $\sim 1.3$ Mpc and is a previously
undiscovered Giant Radio Galaxy \citep[GRG;][]{Schoenmakers2001,
  Saripalli2005}. Automated methods such as the likelihood ratio technique in this case would be able to match the
core of the radio source to an optical counterpart, but would
incorrectly match the lobes to some other optical source, if considered separately.   

\subsection{Positional Offsets}

\begin{figure}
\includegraphics*[width=0.47\textwidth]{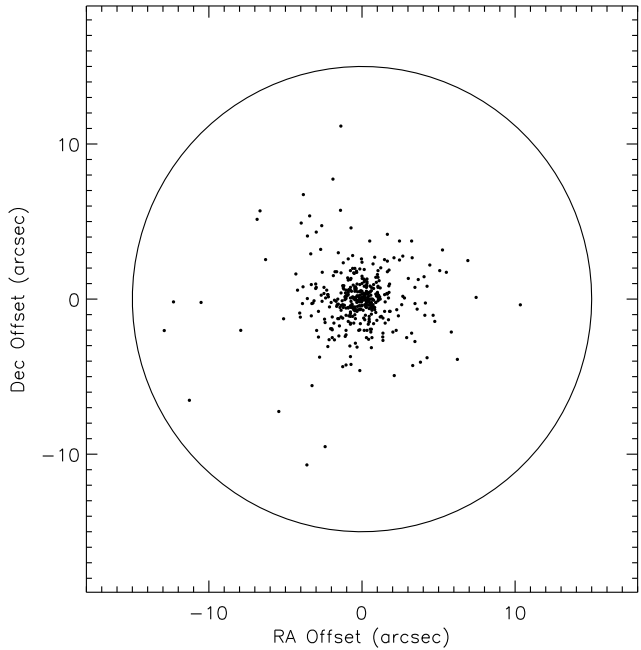}
\caption{The positional offsets between the GAMA and GMRT for the 414 single
  component matches out to z =0.6. The circular line represents a radius of $15''$.}
\label{OFFSETS}
\end{figure}

Figure~\ref{OFFSETS} shows the positional offsets between the
GAMA and GMRT positions of the $414$ single component sources. The median
positional offset between the coordinates is found to be $1.51^{\prime\prime}$.    

\begin{figure}
\includegraphics*[width=0.47\textwidth]{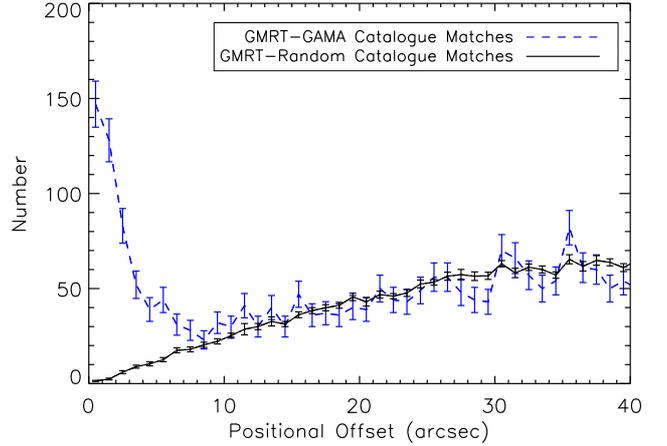}
\caption{Histogram displaying the positional offset between each of
  the $185,125$ GAMA sources in the spectroscopic catalogue, with $z < 0.6$
and $nQ \ge 3$ and $r \le 19.8$, matched to the nearest of the $4,931$ $5\sigma$ GMRT
sources within $3'$ (blue dashed line). This is compared to the positional offsets
between the mean of $10$ random GAMA catalogues matched to the nearest GMRT source (black solid line).}
\label{OFFSET2}
\end{figure}

To test our matching process, in Figure~\ref{OFFSET2}
we show the distribution of the positional offsets between each of the $185,125$ GAMA
sources and the $4,931$ GMRT sources (blue dashed line). Here, this is compared to the positional offsets
between the mean of $10$ GAMA catalogues with randomised positions and
again matched to the nearest GMRT source (black solid line).
The curves converge at a separation of $15''$, which implies that genuine GMRT/GAMA matches should all lie
with $15''$ of one another.

Integrating both curves out to $15''$, yields $778$ matches between the real GAMA/GMRT catalogues
and $267$ matches between the random/GMRT catalogue which indicates
$511$ genuine GAMA/GMRT matches
are expected, which is consistent with our final sample of $499$ ($414$ single and $84$ multiple component) 
GAMA/GMRT matches.

We find that $\sim 10$ per cent ($499$/$4931$) of the radio sources
are matched to GAMA. This is a higher fraction than other matched
radio/optical samples with similar radio flux density limits, for
example \cite{Sadler2002} and \cite{Mauch2007} find that $\sim 1$--$2$
per cent of NVSS radio sources are detected in the 2dFGRS and 6 degree Field
Galaxy Survey \citep[6dFGS][]{Jones2004}. This is due to the fainter
optical limits of GAMA, which probes a larger range of redshifts.

\section{Source Classification}

\begin{figure*}
\subfloat[AGN with pure absorption features.]{\includegraphics[width=0.9\textwidth]{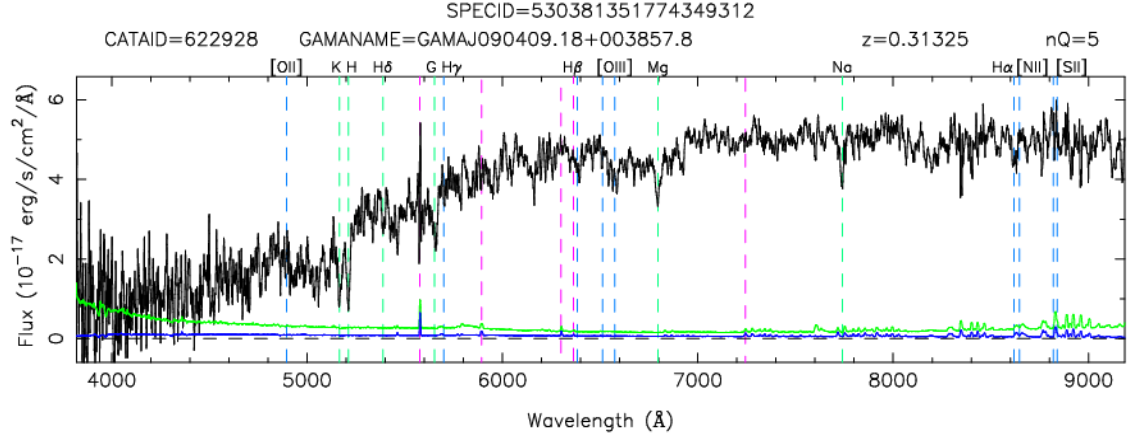}} \\
\subfloat[AGN with broadened emission lines.]{\includegraphics[width=0.9\textwidth]{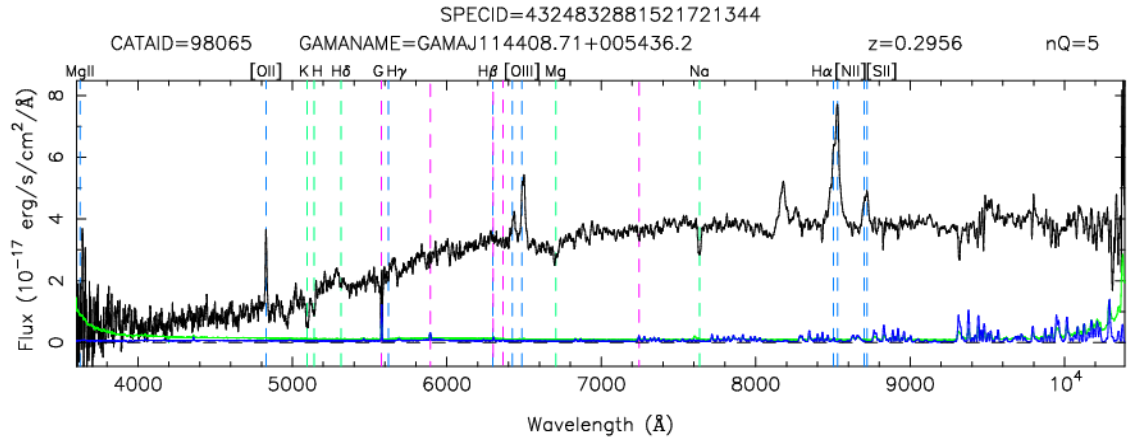}} \\
\subfloat[Star forming galaxy.]{\includegraphics[width=0.9\textwidth]{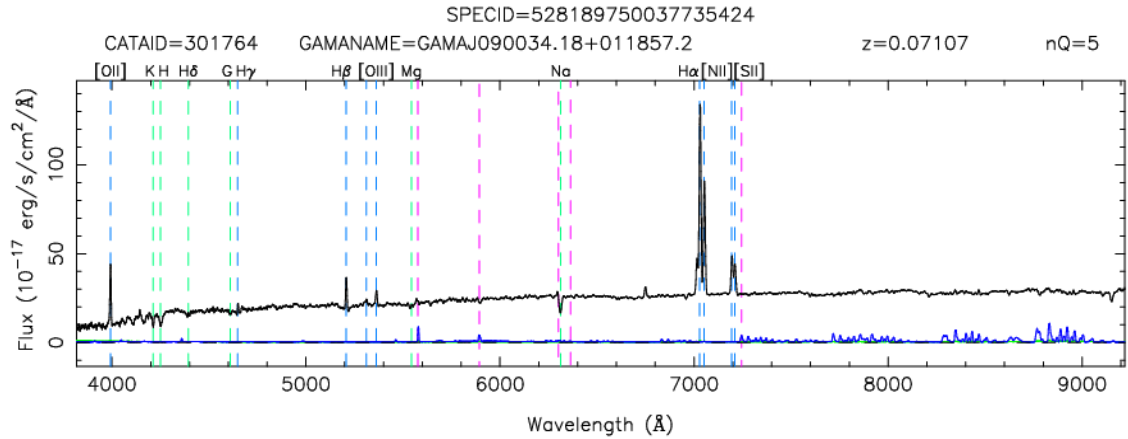}} \\
\caption{Example optical spectra found in our GAMA/GMRT matched
  sample (black lines). Also plotted is the 1 sigma error (green lines) and sky (blue). AGN are classified as having either pure absorption
  features (a) or broadened emission features (b). A
  spectrum of a typical star forming galaxy that have strong, narrow,
  emission lines such as H$\alpha$ can be seen in (c).}
\label{EXAMPLESPEC}
\end{figure*}

After matching the optical/radio counterparts, we divided our sample into
AGN and star-forming (SF) galaxies, by inspecting their individual optical spectra, in a similar
fashion to \cite{Sadler2002} and \cite{Mauch2007}. AGN can have
spectra with either pure absorption features like that of an
elliptical galaxy, broadened emission features (known as Type I AGN),
strong nebular emission [O{\sc ii}], [O{\sc iii}] and [N{\sc ii}] lines compared to the
Balmer series (known as Type II AGN), or spectra with absorption lines
with weak narrow emission lines (LINERS). In this work, sources with optical spectra
revealing strong narrow emission lines consistent with H{\sc ii}
regions, were classified as being star-forming galaxies. Examples of
spectra for AGN and star-forming galaxies can be seen in Figure~\ref{EXAMPLESPEC}. 
The spectrum for each object was inspected independently by three people to ensure that each
classification was reliable and subsequently double-checked with the AGN/SF \cite{Kewley2001} classifications, if it was available in GAMA's emission line
catalogue.     

From the sample of $499$ GAMA/GMRT matches out to $z = 0.6$, we classified 453 sources as AGN and 45 as star-forming 
galaxies. Upon further inspection, one source, GAMA
J142831.94+014331.4, was found serendipitously to be an F5 type radio star in the catalogue of
\cite{Kimball2009}. 

\section{Sample properties}

Each source in our GAMA/GMRT catalogue was detected by the NVSS and we
assigned an NVSS identifcation (and $1.4$ GHz flux density)
during the visual cross-matching process.
In order to determine radio luminosities of each object, we
k-correct the observed flux densities assuming $S_{\nu} \propto
\nu^{\alpha}$ and use spectral indices
determined from the ratios between the 325 MHz and 1.4 GHz fluxes
retrieved from the NVSS catalogue.     

\begin{figure}
\includegraphics[width=0.47\textwidth]{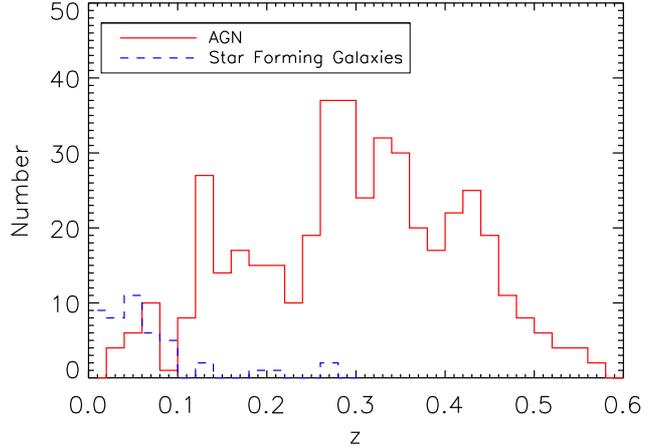}
\caption{Redshift distribution of GAMA-GMRT
  matched AGN (solid red line)
 and star forming galaxies (dashed blue line). The median redshift for
the AGN is $z =0.36$, and $0.06$ for the star forming galaxies.}
\label{Fig2}
\end{figure}

\begin{figure}
\includegraphics[width=0.47\textwidth]{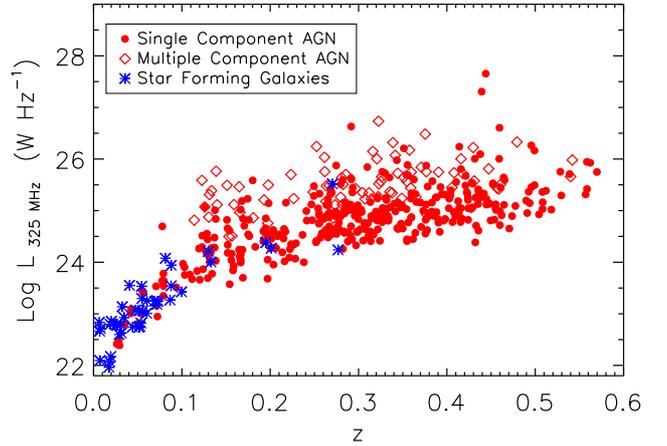}
\caption{The luminosity-redshift distribution of the matched GAMA-GMRT
  sources, showing single component AGN (red filled circles), multiple
  component AGN (red diamonds), and star forming galaxies (blue
  stars). Star forming galaxies become the dominant radio sources at
  $z < 0.1$ at the limits of the GMRT survey.} 
\label{Fig3}
\end{figure}

Median redshifts were found to be $z = 0.36$ for the AGN, and $z=0.06$ for the star forming
galaxies. In Figure~\ref{Fig2} and Figure~\ref{Fig3} we show the
redshift distribution histogram and the
redshift versus radio luminosity plane respectively, for both the AGN and star-forming
galaxies. These figures illustrate that star forming galaxies become the
dominant population of radio sources in the local Universe ($z < 0.1$)
at the limit of $\sim 10$ mJy at 325 MHz.

\subsection{Spectral Indices}

\begin{figure}
\includegraphics[width=0.47\textwidth]{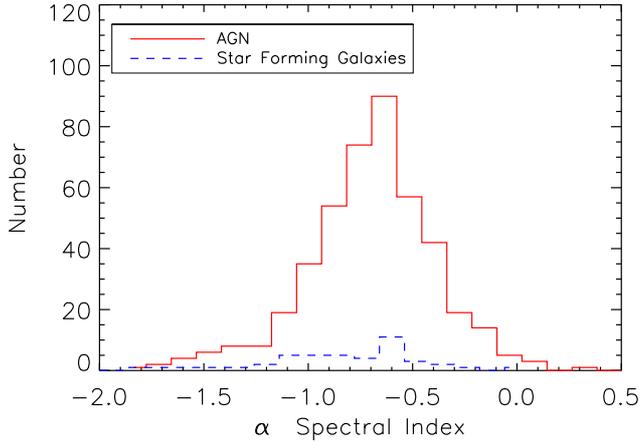}
\caption{The spectral index distribution for AGN (red, solid line) and star forming galaxies (blue, dashed
  line). The spectral indices of the sources are calculated from the ratios
  between 325 MHz GMRT and 1.4 GHz NVSS flux densities.}
\label{SpectralInd}
\end{figure}

Given that every $325$ MHz GMRT source has a detection at $1.4$ GHz
from the NVSS, we can investigate the spectral index distribution for both star-forming galaxies and AGN.
Figure~\ref{SpectralInd} shows the distribution of the spectral
indices across the whole sample. We find that the median spectral indices for AGN and star
forming galaxies
are $\alpha = -0.68 \pm 0.22$  and
$\alpha = -0.82 \pm 0.28$ respectively, which agree well with values
reported by \cite{Mauch2003}, \cite{Owen2009}, \cite{Mauch2013},
\cite{Randall2012}, \cite{Smol2014} and \cite{Coppejans2015}.

\begin{figure}
\includegraphics[width=0.47\textwidth]{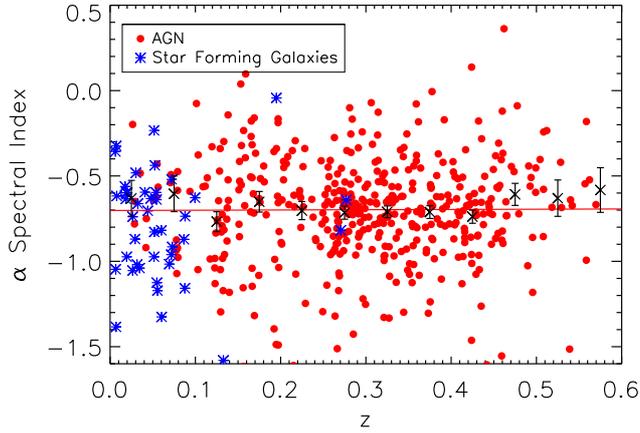}
\caption{The spectral index distribution for AGN (red dots) and star forming galaxies (blue stars) as a function of redshift. The black crosses and error bars indicate the
  mean and standard error on the mean of the spectral index for the AGN in bins of
  $\Delta z = 0.05$. The red line shows the line of
  best fit for the AGN datapoints. The average spectral indices remains more or less
  constant with increasing redshift.}
\label{SpecZ}
\end{figure}

Figure~\ref{SpecZ} shows how the spectral indices of AGN and star
forming galaxies vary as a
function of redshift. In the past, steep spectral indices have been
used to select high-redshift radio galaxy candidates from
low-frequency radio surveys
\citep[e.g.][]{Jarvis2001,DeBreuck2001,Cruz2006}, however we find that there is no significant steepening of the
spectral index in our sample with redshift, albeit over a relatively
low redshift range. This trend was also found by
\cite{Smol2014} out to $z =2.5$, in a 325 MHz VLA survey of the COSMOS
field.    

\begin{figure}
\includegraphics[width=0.47\textwidth]{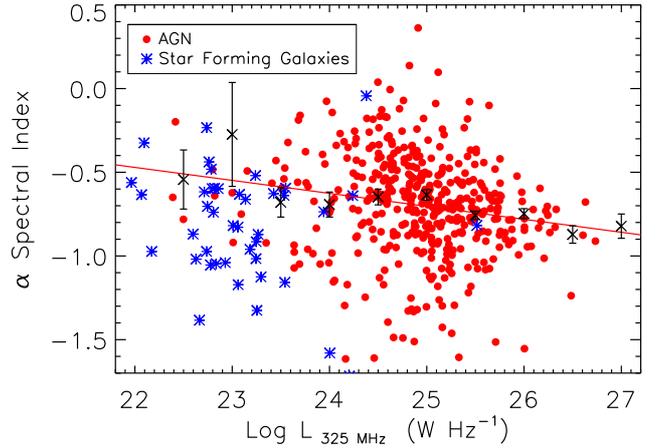}
\caption{The spectral index distribution for AGN (red dots) and star forming galaxies (blue stars) as a function of luminosity. The red solid line shows the line of
  best fit for the AGN datapoints. The black crosses and error bars indicate the
  mean and standard error on the mean of the spectral index for the AGN in bins of
  $\Delta \log_{10} (L_{325}) $ W~Hz$^{-1} = 1$. There is a slight
  decrease in the spectral index with increasing luminosity.}
\label{SPECLUM}
\end{figure}

 In Figure~\ref{SPECLUM}, we show the spectral indices as a function of luminosity. Although there is large scatter in the data,
 we find there is a slight decrease in spectral index with increasing
 luminosity for the AGN. This decrease may arise due to an increasing contribution from higher
luminosity FR II sources whose emission will be dominated by
steep-spectrum optically-thin lobes. A
straight line of best fit to the AGN datapoints yields a gradient of
$m = -0.077 \pm 0.022$ and constant of $c =1.23 \pm 0.55$. A Spearman
Rank correlation test gives a coefficient of $\rho = -0.229$ with a
significance of the deviation from zero of $1.02 \times 10^{-6}$,
indicating a weak but significant correlation. 

\section{Radio Luminosity Functions}

\subsection{Measuring the Radio Luminosity Function} 
To measure the cosmic evolution of the radio sources, we determine the radio
luminosity functions (RLFs) for different redshift slices using the
$1/V_{\rm max}$ method, where $V_{\rm max}$ is the maximum comoving volume
within which a galaxy could lie within a redshift slice and within the
flux limits of the survey \citep{Schmidt1968}. The RLF for a given
luminosity bin is given by:

\begin{equation}
\Phi_{z}(L) = \sum_{i=1}^{N} \frac{1}{C_{i} \times V_{\rm max, i}} ,
\label{EQ1}
\end{equation}

Here $C_{i}$ is the completeness of the survey. We determine upper and lower $1\sigma$ Poisson confidence limits using the
approximation prescribed in \cite{Gehrels1986}.  
As our sample is a matched radio/optical sample we have to take into account
both the optical and radio limits of the surveys, where
$V_{\rm max}$ is the taken as the minimum from the optical and radio $V_{\rm max}$ for each source:
 
\begin{equation}
V_{\rm max} = {\rm min}(V_{\rm max, radio}, V_{\rm max, optical})
\end{equation} 

\begin{figure}
\includegraphics[width=0.47\textwidth]{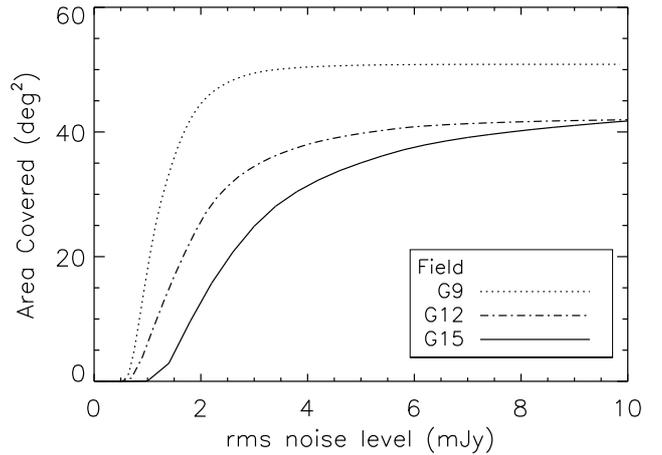}
\caption{Figure showing the cumulative area as a function of rms
  noise level, for each of the three GAMA regions. The total
  areal coverages for G09, G12 and G15 are 50.9,
  42.3 and 45.0 deg$^{2}$ respectively.}
\label{RMSAREA}
\end{figure}

\noindent $V_{\rm max, optical}$ was computed from $z_{\rm max,optical}$, the maximum redshift in which a galaxy can be observed, calculated via SED fitting of the optical spectra described
fully in \cite{Taylor2011}. As the GMRT mosaics have non-uniform
sensitivity, the effective area of the survey changes as a function of
the flux limit. The volume of space available to a source
of a given luminosity $L(V_{\rm max,radio})$ has to be calculated by
taking into account the variation of survey area as a function of flux
density limit. Figure~\ref{RMSAREA} shows how the cumulative area of the survey
varies as a function of rms noise for each of the GMRT mosaics
covering G09, G12 and G15. We use the same method of calculating
$V_{\rm max, radio}$ described in \cite{Smol2009}. Firstly the
differential area in the survey in small bins of rms noise is
determined for each field from Figure~\ref{RMSAREA}. To calculate
$V_{\rm max, radio}$ for a radio source with luminosity $L$, we
compute the maximum redshift ($z_{\rm max,i}$) that the source would have
in each differential rms bin $i$ across the 3 survey fields. $z_{\rm
  max,i}$ is the redshift the source would have before it drops out of
the survey at 5 times the rms of the bin $i$. $V_{\rm max,radio}$ for the
source is then just the sum of all of the individual $V_{\rm max,radio,i}$ normalised by the survey area of that rms bin $A_i$:
\begin{equation}
      V_{\rm max,radio} = \sum\limits_{i=1}^{n} F_i \times
      V_{\rm max,radio,i}(z_{\rm max,i}) ,
\end{equation}
where $F_{i}$ is the fraction of sky corresponding to the area in the
bin $i$, and the sum is over all bins in each of the three fields.

The completeness, $C_{i}$, in Equation~\ref{EQ1} for the RLFs is a combination of the
redshift success and a radio completeness ($C_{i} =
C_{\rm{redshift\,success}} \times C_{R}$). The redshift success is defined as the number of sources with a reliable redshift determination ($NQ \ge 3$) divided by the
total number of sources in the input catalogue within a given r-band
fiber magnitude. In \cite{Loveday2015}, the redshift success is shown
to be well fitted by a modified sigmoid function (see their Equation
1) which we apply here.  
The GAMA survey is highly complete and has a
redshift success of 99 per cent at $r_{\rm Petro} = 19.2$ and 96 per cent at the survey
limit of  $r_{\rm Petro} = 19.8$ (see \cite{Liske2015} for more details).  

\begin{figure}
\includegraphics[width=0.47\textwidth]{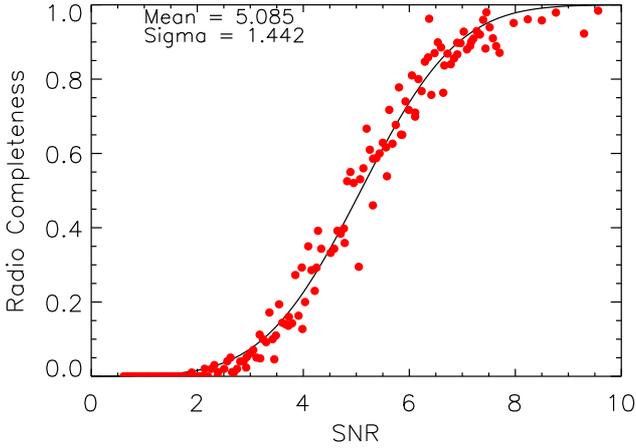}
\caption{Radio completeness as a function of the signal-to-noise
  ratio. Red data points show the fraction of simulated sources
  detected in the GMRT images fitted with a cumulative distribution
  function (black line). In this study we only use sources with an
  SNR$\ge 5$, and so $C_R > 0.5$ for each source.}
\label{RADIOCOMP}
\end{figure}

Errors in fitting fainter sources in the GMRT catalogue can cause it
to be incomplete near the radio survey limit.
We have derived the completeness as a function of signal-to-noise for
the GMRT catalogue by inserting 6000 simulated sources over a range of
flux densities at random positions into 3 images from the survey, one
from each of the G09, G12 and G15 fields. We then used the same source
finding procedure outlined in Section 4 of \cite{Mauch2013} to detect the simulated sources. Figure~\ref{RADIOCOMP} shows the fraction of simulated sources detected in the GMRT images (i.e. the completeness) as a function signal-to-noise ratio. We have fitted the data with the cumulative distribution function:
\begin{equation}
C_{R} = 0.5\left[1 + {\rm erf}\left(\frac{{\rm SNR} - \mu}{\sqrt{2}\sigma}\right)\right]
\end{equation}
Where $C_{R}$ is the radio completeness correction, SNR is the local signal-to-noise ratio, $\mu = 5.1$ and $\sigma=1.4$.

\subsection{Star-forming galaxies}

\begin{figure}
\includegraphics[width=0.47\textwidth]{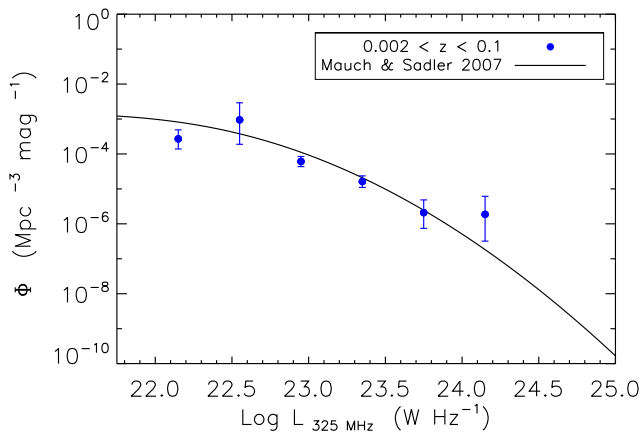}
\caption{The $325$ MHz radio luminosity function for star forming in the
  redshift range $0.002 < z < 0.1$ shown in blue. This is compared to the
  local RLF of \citet{Mauch2007}, which has converted to $325$ MHz from
  $1.4$ GHz assuming a spectral index of $\alpha = -0.7$, shown in
  black. Error bars are determined using the prescription of \citet{Gehrels1986}.}
\label{SFLF1}
\end{figure}

The $325$ MHz radio luminosity function for star-forming galaxies between
$0.002 < z  < 0.1$ for our sample can be seen in Figure~\ref{SFLF1} (and Table~\ref{TableSFValues}). The sample is truncated at $z < 0.1$ to
minimise the effects of evolution. The shape of the RLF is broadly
consistent with the local RLF of \cite{Mauch2007} converted to $325$
MHz from $1.4$ GHz assuming a spectral index of $\alpha = -0.7$.
One key difference between these surveys is the short baseline coverage that allows
extended sources to be recovered in the study of
\cite{Mauch2007}, whereas such sources may be missed in the GMRT
survey. We therefore may not be sensitive to such extended sources,
and our measured RLF should be treated as a lower limit.

Assuming synchrotron self-absorption becomes important in star forming
galaxies, at lower frequencies we would expect a population of flatter spectrum
sources to shift the median spectral index above $\alpha = -0.7$. In
Figure~\ref{SPECLUM} we see no significant population of flat spectrum
sources in the star forming population, which, assuming sychrotron
self absorption means we are likely to be missing sources in the flux
density range of the SF galaxies. These may be missing due to
resolution bias; i.e. they have a peak flux density below the survey
limit, but their total flux density would be above the limit, if they
were detected. This may even be occuring at the relatively
low-resolution and surface brightness sensitivity of the GMRT data.

No evolution measurement is possible from our data, as only six star forming galaxies are detected above $z = 0.1$, with
the majority actually at $z< 0.1$ (Figure~\ref{Fig2}). It is evident we are missing the faintest star forming
galaxies with $\log_{10} L_{325} < 22.0$. To rectify this, deeper
radio data with $\mu$Jy flux limits, over the same area of sky, would
be needed to further constrain the faint-end of the luminosity function at frequencies of 325 MHz. Indeed this is one of the key
science cases for current and future deep radio surveys covering much
smaller areas \citep{Jarvis2014a,McAlpine2015}.

\begin{table}
\centering
\caption{The $325$ MHz Radio Luminosity Function for star forming galaxies.}
\label{TableSFValues}
\begin{tabular}{cccc}
\hline
 Luminosity & Number Density & Number \\
 ($\log_{10} (L_{325-\rm{MHz}})$)  & (${\rm Mpc^{-3} mag^{-1}}$) & \\
\hline
$22.15$  &\T\B   $2.72^{+2.22}_{-1.32} \times 10^{-4}$  &  $4$ \\    
$22.55$  &\T\B   $1.00^{+2.11}_{-0.81}  \times 10^{-3}$  &  $7$ \\  
$22.95$  &\T\B  $6.10^{+2.36}_{-1.76} \times 10^{-5}$  &  $14$ \\
$23.35$  &\T\B  $1.63^{+0.73}_{-0.53}  \times 10^{-5}$  &  $10$ \\
$23.75$  &\T\B   $2.09^{+2.75}_{-1.35}  \times 10^{-6}$  &  $2$ \\
$24.15$  &\T\B  $1.86^{+4.26}_{-1.54} \times 10^{-6}$  &  $1$ \\
\hline 
Total & &  38 \\
\hline
\end{tabular}
\end{table} 

With these caveats in mind we can derive a lower limit of the local
star-formation rate density, using the 1.4 GHz luminosities of the galaxies and converting them to star formation
rates (in M$_{\odot}$ yr$^{-1}$) via the relation presented in
\cite{Bell2003}, we estimate the total star formation rate density as $
\rho_{\rm SFR} =\Sigma ({\rm SFR}/V_{\rm max})  = 0.013  \pm 0.002$ M$_\odot$
  yr$^{-1}$ Mpc$^{-3}$. 

Many different estimators have been used to determine the star
formation rate density of the Universe. At 1.4 GHz \cite{Mauch2007} find the star formation rate of the local Universe
  to be $\rho_{\rm SFR} = 0.022 \pm 0.001$ M$_{\odot}$ yr$^{-1}$
  Mpc$^{-3}$, which is slightly greater than our result. Using H$\alpha$ measurements \cite{Gunawardhana2013}
have also estimated the evolution of the star formation density out to
$z < 0.34$. Using values from their Table 3, $\rho_{\rm SFR} =  0.024
\pm 0.006$ M$_{\odot}$ yr$^{-1}$ Mpc$^{-3}$ between $0 < z <
0.24$. \cite{Westra2010} using data from the Smithsonian Hectospec Lensing
Survey (SHELS) measured a value of $\rho_{\rm SFR} = 0.018$
M$_{\odot}$ yr$^{-1}$ Mpc$^{-3}$ out to $z < 0.2$. \cite{James2008}
found the star formation density of local star forming galaxies to lie
between $\rho_{\rm SFR} = 0.016$ and $\rho_{\rm SFR} = 0.023$
M$_{\odot}$ yr$^{-1}$ Mpc$^{-3}$ from a sample of galaxies in the
H$\alpha$ Galaxy Survey.     

\subsection{AGN} 

\begin{figure*}
\includegraphics[width=0.47\textwidth]{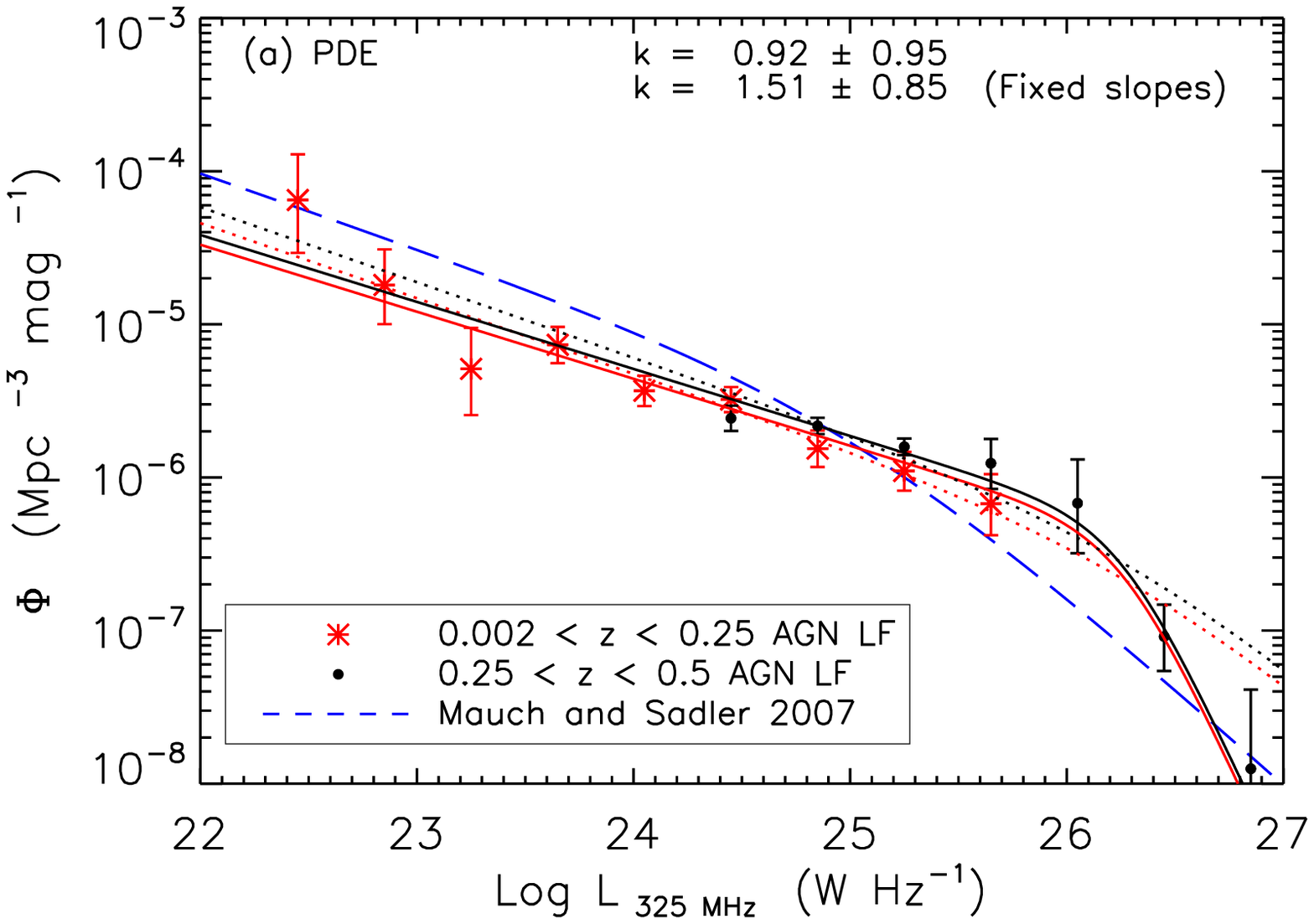}
\includegraphics[width=0.47\textwidth]{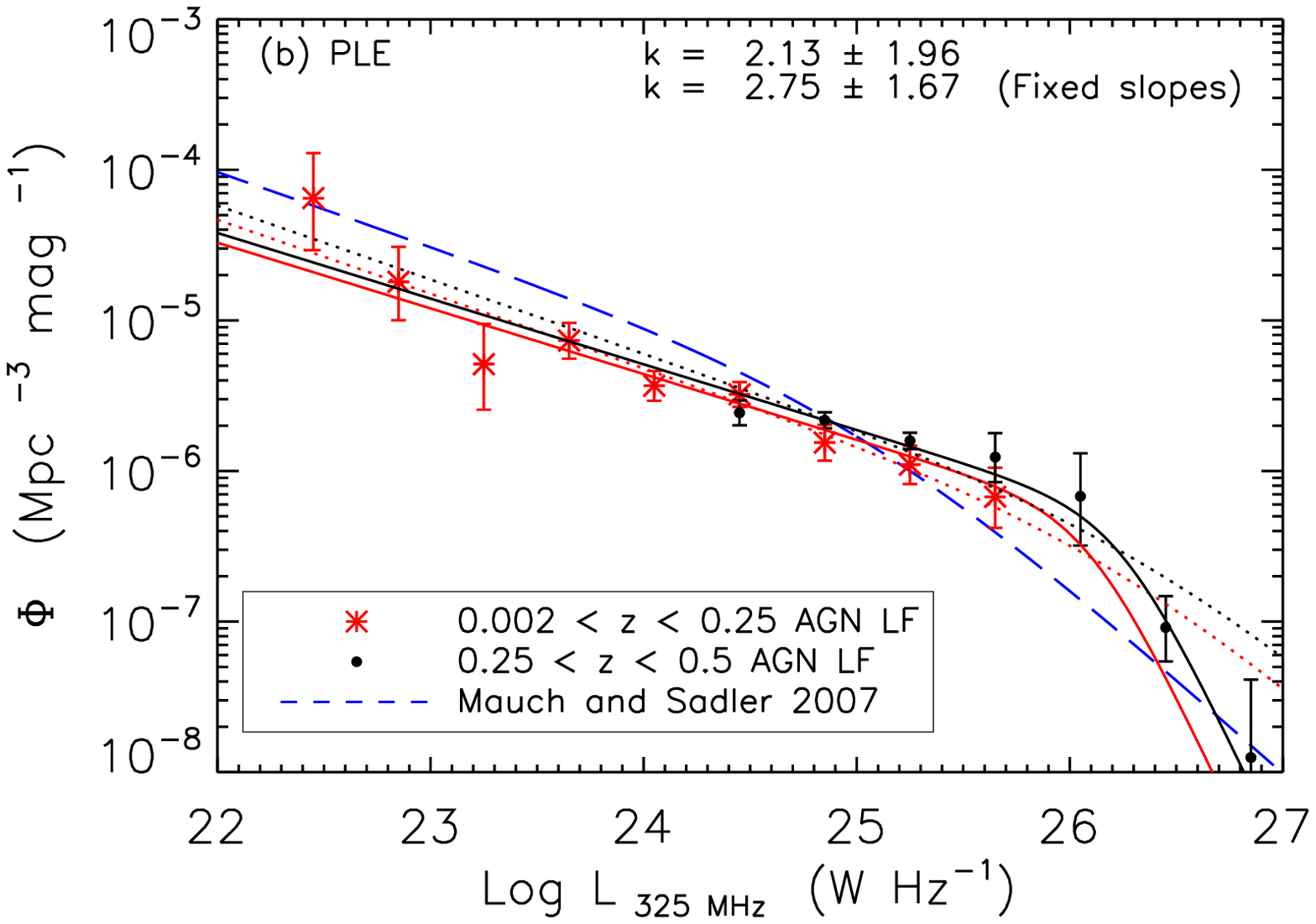}
\caption{$325$ MHz AGN radio luminosity functions for the
  redshift ranges $0.002 < z < 0.25$ (red crosses) and $0.25 < z <
  0.5$ (black filled circles). Panels (a) and (b) shows the best fitting RLF assuming
  pure density and pure luminosity evolution respectively (solid lines).    
The best fitting RLFs found by fixing the bright and
faint-end slopes to be equal to those of \citet{Mauch2007} can be seen
as dotted lines. The local RLF of \citet{Mauch2007}, which has been converted to 325 MHz from
  1.4 GHz assuming a spectral index of $\alpha = -0.7$ can be seen as
  a blue dashed line. Bin sizes of  $\Delta \log_{10} L = 0.4$ are
  used. Error bars are determined using the prescription of \citet{Gehrels1986}.}
\label{AGNEvolve}
\end{figure*}

The RLFs for AGN in two redshift slices ($0.002 < z < 0.25$ and $0.25
< z < 0.5$) are shown in Figure~\ref{AGNEvolve} (See Table~\ref{TableAGN} for the RLF values). We find that the 325 MHz
RLF at these redshifts is comparable to the local 1.4 GHz RLF of
\citet{Mauch2007} converted to 325 MHz assuming a spectral index of
$\alpha = -0.7$. There also appears to be evidence for positive
evolution, at least at $L_{325} > 10^{25}$~W~Hz$^{-1}$.

In order to quantify the cosmic evolution of the AGN,  following
\citet{Dunlop1990}, we fit a double
power law function to the data given by:

\begin{equation}
 \Phi(L) = \frac{C}{(L_{*}/L)^{A} + (L_{*}/L)^{B}} ,
\end{equation} 

\noindent where $C$ is the normalisation, $L_{*}$ is the 
luminosity corresponding to the break in the RLF and $A$ and $B$ are the bright and faint
end slopes. We simultaneously fit both redshift slices for evolution assuming two
scenarios for the RLF, one in which the luminosity of the radio sources is
fixed and undergoes pure density evolution parametrized by:

\begin{equation}
\Phi_{z}(L) = (1+z)^{k}\Phi_{0}(L),
\end{equation}

\noindent and another in which the number density of radio sources is
fixed and the population undergoes pure luminosity evolution: 

\begin{equation}
\Phi_{z}(L) =\Phi_{0}\Bigg(\frac{L}{(1+z)^{k}}\Bigg),
\end{equation}

\noindent where $\Phi_{z}(L)$ is the RLF at redshift $z$,
$\Phi_{0}(L)$ the normalisation of the 
local RLF, and $k$ denotes the strength of the evolution. 

\begin{table*}
\centering
\caption{The $325$ MHz Radio Luminosity Function for AGN.}
\label{TableAGN}
\begin{tabular}{ccccc}
\hline
& \multicolumn{2}{c}{$0.002 < z < 0.25$} & \multicolumn{2}{c}{$0.25 < z < 0.5$} \\
 Luminosity & Number Density & Number  & Number Density & Number \\
 ($\log_{10} (L_{325-\rm{MHz}})$)  & (${\rm Mpc^{-3}
   mag^{-1}}$) & &
 (${\rm Mpc^{-3} mag^{-1}}$) &\\
\hline
$22.4$  &\T\B $6.49^{+6.41}_{-3.56} \times  10^{-5}$  &  $3$ &  &\\
$22.8$  &\T\B  $1.81^{+1.28}_{-0.80} \times 10^{-5}$  &  $5$ &  & \\ 
$23.2$  &\T\B  $5.12^{+4.36}_{-0.26}  \times 10^{-6}$  &  $4$ &  &  \\
$23.6$  &\T\B  $7.35^{+2.27}_{-1.78}  \times 10^{-6}$  &  $19$ &  & \\
$24.0$  &\T\B  $3.68^{+0.93}_{-0.76}  \times 10^{-6}$  &  $26$ &  & \\
$24.4$  &\T\B  $3.22^{+0.67}_{-0.56}  \times 10^{-6}$ &  $35$ & $2.44^{+0.50}_{-0.42}  \times 10^{-6}$ &$45$\\
$24.8$  &\T\B  $1.54^{+0.48}_{-0.37}  \times 10^{-6}$ & $19$ & $2.17^{+0.28}_{-0.25}  \times 10^{-6}$  &$93$\\
$25.2$  &\T\B  $1.10^{+0.36}_{-0.28}  \times 10^{-6}$ & $15$ & $1.59^{+0.21}_{0.19}  \times 10^{-6}$  &$83$\\
$25.6$  &\T\B  $6.74^{+3.77}_{-2.55}  \times 10^{-7}$ & $8$ & $1.24^{+0.55}_{-0.40}  \times 10^{-6}$  &$47$\\
$26.0$  &\T\B   &  & $6.79^{+6.32}_{-3.59} \times 10^{-7}$ &$18$\\
$26.4$  &\T\B   &  & $9.13^{+5.62}_{-3.69}  \times 10^{-8}$  &$6$\\
$26.8$  &\T\B   &  & $1.25^{+2.88}_{-1.04} \times 10^{-8}$ &$1$\\
$27.2$  &\T\B   & & $1.25^{+2.88}_{-1.04} \times 10^{-8}$ &$1$\\ 
\hline 
Total & & $134$ & & $294$\\
\hline
\end{tabular}
\end{table*} 
 
Figure~\ref{AGNEvolve} shows the best fits for both evolution
scenarios. We fit these assuming both a fixed and variable bright and faint-end slopes of
the RLF. For the fixed slopes, we use those
found in the local 1.4 GHz luminosity function of
\cite{Mauch2007}, with $A = 1.27$ and $B=0.49$.
 The parameter values obtained from the fits are shown in Table~\ref{Table2}. 

We find evidence of mild evolution in the AGN RLF. Evolution parameters of $k = 0.92 \pm 0.95$ in the case of
pure density evolution and $k = 2.13 \pm 1.96$ for pure luminosity
evolution, are obtained without fixing the faint and bright-end slopes. 
Fixing the slopes to the values found by \cite{Mauch2007},  we find
$k = 1.51 \pm 0.92$ for pure density evolution and $k = 2.75 \pm 1.51$ for
pure luminosity evolution. The values of the bright-end slopes we find when fitting
for all the parameters are steeper than those found by
\cite{Mauch2007}, but the faint-end slope is very similar.
$L_{*}$ and $C$ are similar in all the fits, with $\log_{10}(L_{*}
\approx 26.25$ W~Hz$^{-1}$) and the normalisation $C \approx -6.5$
Mpc$^{-3}$ in all cases.       

\begin{table*}
\centering
\caption{Best fitting parameters from fitting the 325 MHz RLF for pure luminosity
  and pure density evolution, assuming a double power law as in Dunlop
  et al. (1990). We fit the RLFs with and without fixing the power law slopes
  as those found in Mauch and Sadler. (2007).}
\label{Table2}
\begin{tabular}{ccccc}
\hline
\multicolumn{5}{c}{Evolution Scenario} \\
Parameter  & PDE & PDE (Fixed Slopes) & PLE  &
PLE (Fixed Slopes) \\
\hline 
$\log_{10}(L_{*})$ W~Hz$^{-1}$  & $26.26 \pm 0.15$  & $26.43 \pm 0.51$&$25.96 \pm 0.29$  & $26.11 \pm 0.52$ \\ 
$\log_{10} (C)$ Mpc$^{-3}$     & $-6.40 \pm 0.19$  & $-6.60 \pm 0.27$&$-6.27 \pm 0.15$  &  $-6.43 \pm 0.26$ \\
$A$ & $3.08 \pm 1.62$  & $1.27$ (Fixed)  & $3.02 \pm 1.56$& $1.27$ (Fixed) \\ 
$B$  & $0.44 \pm 0.06$  & $0.49$ (Fixed)  &$0.44 \pm 0.06$& $0.49$ (Fixed) \\ 
$k$         & $0.92 \pm 0.95$  & $1.51 \pm 0.92$  & $2.13 \pm 1.96$  & $2.75 \pm 1.51$ \\ 
\hline
Reduced $\chi^{2}$                               &  $0.80$ & $1.21$ & $0.79$ & $1.22$ \\
\end{tabular}
\end{table*} 

We also fit straight lines of best fit to the AGN data for the whole range
of luminosities and those with $\log_{10}L_{325} < 24.5$ W~Hz$^{-1}$, which have
the form in the case of pure density evolution:

\begin{equation}
\log_{10}\Phi_{z} = m\log_{10}L + k\log_{10}(1+z) + c ,
\end{equation} 

\noindent and in pure luminosity evolution:

\begin{equation}
\log_{10}\Phi_{z} = m(\log_{10}(L) -  k\log_{10}(1+z)) + c .
\end{equation} 

\noindent These fits yield the values shown in
Table~\ref{Table3}. Stronger evolution is found for both scenarios, when fitting the whole range of
luminosities, which indicates that lower luminosity
sources undergo less evolution than
higher luminosity sources $\log_{10}L_{\rm{325-MHz}} > 24.5$ W~Hz$^{-1}$ a trend
which has been seen by \cite{Waddington2001, Willott2001, Tasse2008, Donoso2009, Rigby2011, Simpson2012}. 

\begin{figure}
\includegraphics[width=0.47\textwidth]{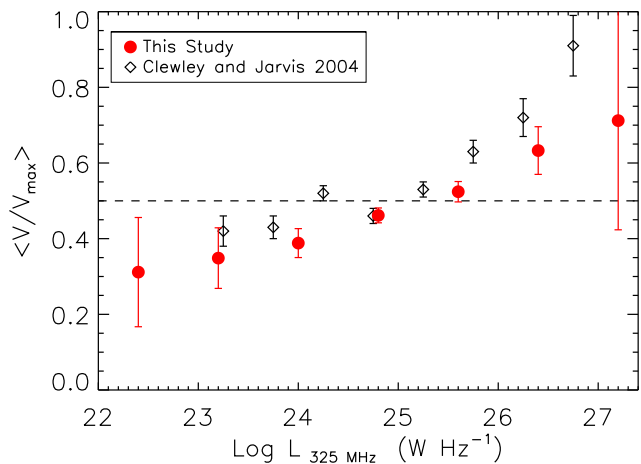}
\caption{ The $V/V_{\rm max}$ statistic as a function
the radio luminosity for AGN out to $z =0.5$ for the GMRT data (red circles), compared to the
results of Clewley and Jarvis 2004 (black diamonds). Errors bars are given as
$1/\sqrt{12N}$, as in \citet{Condon2002}.}
\label{VVmax}
\end{figure}

To investigate this further, we follow \cite{Clewley2004} and
calculate the parameter free $V/V_{\rm max}$ statistic as a function
the radio luminosity. The $V$ in this case is the volume enclosed by
the source at its true redshift and $V_{\rm max}$ is the volume up to
which it could still be found given the survey parameters, i.e. the
same $V_{\rm max}$ used to calculate the RLF. For a non-evolving
population, then the radio sources should be uniformly distributed
between 0 and 1 for this statistic, resulting in a constant co-moving
population giving a value of $V/V_{\rm max} = 0.5$, whereas a
negatively evolving population would give $V/V_{\rm max} < 0.5$ and a
positively evolving population resulting in $V/V_{\rm max} >
0.5$. Figure~\ref{VVmax} shows the  $V/V_{\rm max}$ statistic for our
sample, at radio luminosities $L_{\rm{325 MHz}} < 10^{25}$~W~Hz$^{-1}$ the population does not appear to be
  evolving, whilst at $L_{\rm{325 MHz}} > 10^{25}$~W~Hz$^{-1}$ there appears to be some evidence for positive
  evolution. At the faintest luminosities it is possible that we don't
  see evolution due to the lack of volume probed. This is similar to what was found by \cite{Clewley2004} albeit with a
radio survey with a higher flux limit, and using photometric
redshifts. \cite{Tasse2008} also show there is more positive
  evolution for radio loud AGN with $L_{\rm{1.4-GHz}} >
  10^{24.5}$~W~Hz$^{-1}$ than at lower lumonisities, consistent with
  the results here, using a sample of AGN observed at frequencies
  of $74$, $230$, $325$ and $610$ MHz using the VLA and GMRT in the XMM-LSS field.  

\begin{table*}
\centering
\caption{Straight line of best fit parameters for the 325 MHz RLF
  assuming pure luminosity and pure density evolution for the whole
  range of luminosities and for AGN with $\log_{10}L_{325} < 25.5$ W~Hz$^{-1}$.}
\label{Table3}
\begin{tabular}{ccccc}
\hline
\multicolumn{5}{c}{Evolution Scenario} \\
Parameter  & PDE & PLE &PDE ($\log_{10}L_{325} < 25.5$ W~Hz$^{-1}$) &PLE ($\log_{10}L_{325} < 25.5$ W~Hz$^{-1}$) \\
\hline 
$m$   & $-0.53 \pm 0.05$  & $-0.56 \pm 0.04$  & $-0.44 \pm 0.06$  & $-0.43 \pm 0.06$ \\ 
$c$    & $ 7.28 \pm 1.18 $  &  $8.26 \pm 1.06$   &  $5.21 \pm 1.46$ &  $5.02 \pm 1.43$ \\
$k$  & $1.46 \pm 0.93 $ &  $3.04 \pm 1.48$   &  $0.81 \pm 1.01$   &$1.66 \pm 2.13$\\                               
\hline
Reduced $\chi^{2}$ & $1.53$ & $2.34$ & $0.97$ & $1.03$ \\
\end{tabular}
\end{table*} 

\subsection{High and Low Excitation AGN}
The optical spectra allow us to separate the AGN population into high
and low excitation radio galaxies (HERGs and LERGs). As in \cite{Laing1994} and \cite{Best2012} we make use
of the $5007 \AA$  [OIII] line to divide the HERG and LERG populations. Here we
define HERGs as those as having a measurable [OIII]
equivalent widths (EW) $> 5 \AA$, which results in a sample of 68 HERGs and 382 LERGs.  

\begin{figure}
\includegraphics[width=0.47\textwidth]{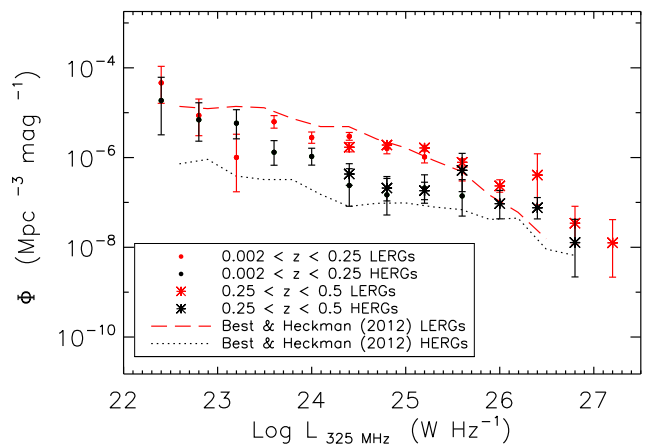}
\caption{Radio Luminosity Functions for high (red datapoints)
  and low (black datapoints) excitation
AGN in two redshift slices $0.002
  < z < 0.25$ (filled circles) and $0.25 < z < 0.5$ (stars). The
  local 1.4 GHz HERG and LERG RLFs (converted to $325$ MHz using a
  spectral indices of $\alpha = -0.7$) of Best and Heckman 2012, can be
  seen as the red dashed and black dotted lines respectively.}
\label{HERGLERG}
\end{figure}

Figure ~\ref{HERGLERG} shows the RLFs for HERGs and LERGs in two
redshift slices $0.002  < z < 0.25$ and $0.25 < z < 0.5$, compared to
the local $1.4$ GHz HERG and LERG RLFs of \cite{Best2012}
converted to 325 MHz, assuming a spectral index of $\alpha = -0.7$.  
A clear division between the HERG and LERG RLFs can be seen, at luminosities between $
10^{23} < L_{\rm{325-MHz}} < 10^{26}$ W~Hz$^{-1}$, and the LERGs are the
dominant population. We calculate that the fraction of HERGs decreases from $0.3$ between $0.002 < z
< 0.25$ to $0.21$ between $0.25 < z < 0.5$. The faint end and bright ends of the RLFs tend to
converge.  Comparing our results with those of \cite{Best2012} we find
that the LERG RLFs agree well with each other, whereas the HERG RLFs tends to diverge at lower
luminosities $L_{\rm{325-MHz}} < 10^{24}$ W~Hz$^{-1}$.    

As our sample size is small we are unable to observe any
evolution. Pracy et al. (submitted) will contain a detailed
investigation of the evolution of HERGs and LERGs at 1.4 GHz, making
use of GAMA redshifts, as well as a discussion of the divergence of
the HERG RLFs at lower luminosities.     

\subsection{Steep and Flat Spectrum AGN}
In the simplest form of the orientation-based unification
of AGN, we expect the flat-spectrum sources to be a subset of
the parent steep-spectrum population that happen to have their radio
jets oriented along our line of sight \citep{Urry1995}. These
flat-spectrum sources would therefore have their radio flux density
boosted due to relativistic beaming effects, and their intrinsic
luminosity could be significantly lower than observed.

\begin{figure}
\includegraphics[width=0.47\textwidth]{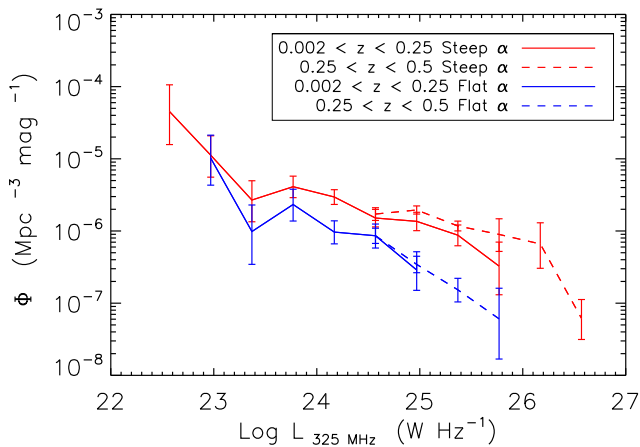}
\caption{Radio Luminosity functions of AGN with steep ($\alpha <
  -0.5$, red) and flat ($\alpha > -0.5$, blue) spectral indices, for two redshift slices $0.002
  < z < 0.25$ (solid lines) and $0.25 < z < 0.5$ (dashed lines)}. 
\label{AGNDIFF}
\end{figure}

One of the unique aspects of the 325 MHz data is that we have spectral
indices for the entire sample of radio-loud AGN, therefore we are able to
measure separable RLFs for AGN with steep ($\alpha < -0.5$) and flat
($\alpha > -0.5$) spectra. 
Figure~\ref{AGNDIFF} shows the RLFs for the
two populations for two different redshift slices; $0.002 < z < 0.25$ and
$0.25  < z < 0.5$. The steep spectrum sources are seen to be more numerous 
and span a wider range of luminosities than the flat spectrum sources
in our sample. 

In order to determine whether the flat-spectrum objects are indeed a subset of the
steep-spectrum sources, or conversely, we assume that they are and derive
beaming parameters, to estimate the shift that is required on the
flat-spectrum RLF for it to sit on the steep-spectrum RLF.

\begin{figure}
\includegraphics[width=0.47\textwidth]{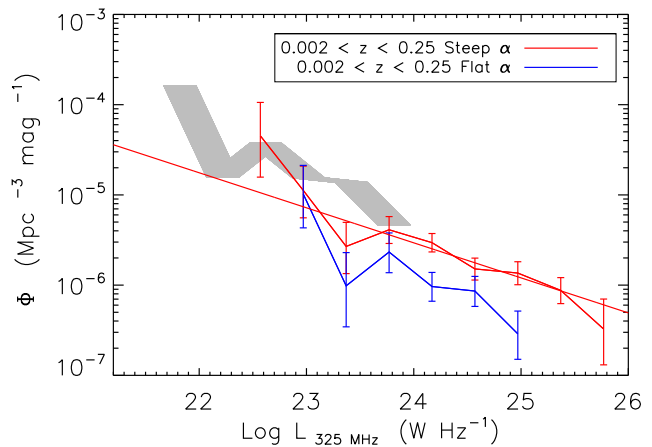}
\caption{Figure showing that the RLF (in the redshift slice $0.002 < z < 0.25$) for flat-spectrum sources
  (blue line) can be shifted in number density and luminosity (grey region), to incide with an extrapolation of
  the steep-spectrum sources (red), as is expected from the orientation based unification of radio-loud AGN.}
\label{FLATSTEEP}
\end{figure}

The radio flux density is enhanced as $\Gamma^2$:
\begin{equation}
\Gamma = \gamma^{-1} (1 - \beta\cos\theta)^{-1},
\end{equation}
where $\gamma$ is the Lorentz factor, $\beta = v/c$ and $\theta$ is
the angle between the radio jet and the line of sight. Following
\cite{Jarvis2002}, we adopt a conservative approach and assume that the flat-spectrum population have an
opening angle of $\theta \sim 20^{\circ}$, which suggests a flux-boosting
factor of the order of 10-20 over the steep-spectrum population. Furthermore, given the opening angle of
$20^{\circ}$ we can calculate the fraction of the parent population we
are observing as flat-spectrum. Assuming a spherically symmetric
system, then an opening angle of 20$^{\circ}$ corresponds to a
fraction of 6 per cent of the parent population, from which we infer
that the measured space density is a factor of $\sim 16$ lower than
it would be if we could see the un-boosted population. To check
if this is consistent with the RLF of the steep-spectrum sources we
shift the flat-spectrum RLF by a factor of 10 in luminosity,
and a factor 16-20 in number density. Figure ~\ref{FLATSTEEP} shows that this extrapolation of
the steep-spectrum RLF (seen in grey) does indeed agree with the shifted RLF of the
flat-spectrum population, as would be expected from orientation-based
unification of radio-loud AGN. However, we have assumed that all of the flux from
the flat-spectrum sources is Doppler boosted, whereas in reality some
flux is undoubtedly  emitted by extended jets or lobes which are not
associated with the core of the radio sources. As such our result that
the flat-spectrum population forms a subset of the steep-spectrum
population is only an indication 
that the orientation-based unification of radio-loud AGN is consistent
with the data, rather than confirmation.
 
\section{Discussion}

The results presented use data from a deep and wide 325 MHz
survey from \cite{Mauch2013}, combined with the optical imaging and
spectroscopic redshifts from the GAMA survey. This sample is the
largest using spectroscopic redshifts obtained to date that spans a significant enough range in
redshift ($0.002 < z <0.5$) to measure evolution.  

 We find that the 325 MHz RLF for star-forming galaxies, at $z < 0.1$, is in agreement with those produced by
combining 1.4 GHz NVSS and optical data from the 6dFGRS
\citep{Mauch2007}. In order to determine a radio derived estimate of the evolution of star formation
rate density from the GAMA survey, we would
require much deeper GMRT data going to lower flux limits.
\cite{Padovani2011} for instance found that star-forming galaxies
evolve as $L \propto (1+z)^{2.8}$, using 1.4 GHz data to flux
limits of $50$ $\mu$Jy from the VLA-CDFS survey.   
 Using 1.4 GHz VLA data matched to the VIDEO survey
\cite{McAlpine2013}, found that the RLF of star-forming galaxies
evolves as $L \propto (1+z)^{2.5}$, which is consistent with many other tracers of star formation
\citep{Hopkins2003, Hopkins2004, Prescott2009}.      

Our measured 325~MHz RLF of AGN is in broad agreement with the 1.4~GHz
RLF measued by \citep{Mauch2007} when converted to 325~MHz using a
spectral index of $\alpha=-0.7$. The \cite{Mauch2007} LF also
  consistent with the local 20 GHz LF of \cite{Sadler2014}, it therefore appears
that the local AGN population is remarkably uniform over a large range in
frequency (325 MHz to 20 GHz). 

 We note that, at the faint end, the
1.4~GHz RLF is consitently higher than our 325~MHz RLF, assuming the
spectral index correction. The increased space density
of 1.4 GHz selected galaxies may be due to doppler boosting of the
population, whereby beamed 1.4~GHz sources are shifted to higher
luminosity. Radio-loud AGN samples selected at lower frequency are not
as likely to detect significant numbers of sources with beamed
cores, as at lower frequency the lobes of galaxies should
dominate. This implies that the 325 MHz LFs presented here are
  not biased with respect to orientation and are more representative
  of the mechanical jet power in the local radio-loud AGN population. 

Regarding evolution, we detect mild but poorly constrained evolution
from fits to the 325 MHz RLF for AGN out to
 $z =0.5$, with evolution parameters of $k = 0.92 \pm 0.95$ for pure density evolution 
and $k = 2.13 \pm 1.96$ for pure luminosity evolution. Considering
just the low-luminosity population (those with $L_{\rm{325-MHz}} <
25.5$ W~Hz$^{-1}$) the evolution appears to be much less.   
This is in line with previous studies (see Table~\ref{Table3}), principally based on smaller
samples from deep fields, where the low-luminosity population evolves
significantly less strongly that the high-luminosity population. 
Our results are also consistent with \cite{Donoso2009} and
  although they did not estimate an evolution parameter, they
  found that the number density of low-luminosity AGN increases by
  a factor of $\sim 1.5$, out to $z =0.55$, which implies an evolution
  parameter of $k \sim 0.93$.     

Low-luminosity sources are thought to be the population that
dominates the mechanical feedback in the $z< 1$ Universe
\citep[e.g.][]{Croton2006,Smol2009, Smol2009b, Best2012, Smol2015}, and thus constraining their evolution is
key to our understanding of the relevance of such feedback in studies
of galaxy evolution. However, we note that some fraction of such
sources are thought to be powered by the efficient accretion of cold gas,
determined by their emission-line properties \citep[e.g.][]{Simpson2012,Best2014,Mingo2014}.

Historically the vast majority of continuum surveys that have produced RLFs have been
conducted at 1.4 GHz. Our result is in agreement with the only other comparable study
 concerned with the evolution of radio sources at 325 MHz by
 \cite{Clewley2004}, who by matching SDSS DR1 data with the Westerbork
 Northern Sky Survey (WENSS), found no evolution in low-luminosity AGN via a $V/V_{\rm max}$ analysis, out to $z =0.8$. 

\begin{table*}
\centering
\caption{Comparison of the evolution parameters for radio-loud AGN determined from previous
  studies discussed above.}
\label{Table3}
\begin{tabular}{ccccccc}
\hline
Study  & Frequency & Flux Limit & $z$ Range & Sample Size &
Evolution Parameter $k$ &Redshifts \\
\hline
This study & 325 MHz  & $\sim 5$ mJy  & $0.0 < z < 0.5$ & $428$ & $ 2.13
\pm 1.96 $ PLE & spectro \\
& & & & & $0.92 \pm 0.95$  PDE \\ 
\cite{Brown2001} & 1.4 GHz & $5$ mJy & $0.0 < z < 0.4$ & $230$ & $3 - 5$ PLE & photo\\ 
\cite{Sadler2007} & 1.4 GHz & $2.8$ mJy  & $0.0 <  z < 0.7$ & $391$ & $2.3
\pm 0.3$ PLE & spectro\\ 
\cite{Smol2009} & 1.4 GHz & $50$ $\mu$Jy & $0.0 < z < 1.3$ & $601$ &
$2.0 \pm 0.3$ PLE & combined\\
\cite{McAlpine2011} & 1.4 GHz & $1$ mJy & $0.0 <
z<0.8$& $131$ & $0.8 \pm 0.2$ PLE & photo\\
 &  & & &  & $0.6 \pm 0.1$ PDE & \\
\cite{Padovani2011} & 1.4 GHz & $50$ $\mu$Jy & $0.1 <
z < 5.8$& $86$ & $-3.0 \pm 1.0$ PLE & combined \\
 &  &  & &  & $-1.6 \pm 0.4$ PDE & \\
\cite{McAlpine2013} & 1.4 GHz & $100  \mu$Jy  & $0 < z < 2.5$  & $951$
& $1.18 \pm 0.21$ PLE & photo\\
\cite{Padovani2015} & 1.4 GHz & $50$ $\mu$Jy & $0.1 <
z < 4.5$& $136$ & $-6.0 \pm 1.4$ PLE & combined \\
&  &  & &  & $-2.4 \pm 0.3$ PDE & \\

\hline 

\end{tabular}
\end{table*}

\section{Conclusions}

We have produced $1/V_{\rm max}$ radio luminosity functions for samples of AGN and star forming
galaxies out to $z \sim 0.5$, by combining data from the largest survey conducted to date at
325 MHz from the GMRT, with the GAMA survey. Our main results are as follows:  

\begin{enumerate}

\item By cross-matching a 325 MHz GMRT survey covering
  $138$ deg$^{2}$ of the GAMA spectroscopic survey, we are able to
  produce a GMRT/GAMA matched sample of $499$ objects with $z \le 0.6$, $nQ \ge 3$ and $r \le
  19.8$. Inspection of the optical spectra in our sample allows us to
  divide the sample into $45$ star forming galaxies and $453$ AGN.  

\item  The mean spectral index of AGN remains constant with redshift as found in \cite{Smol2014}
  (Fig.~\ref{SpecZ}). Higher luminosity AGN are found to have slightly steeper spectral
  indices than those with lower luminosities possibly due to the
  increase in fraction of FR II sources (Fig.~\ref{SPECLUM}).   

\item We determine the local luminosity function of star forming
  galaxies at 325 MHz, which is broadly consistent with the local 1.4 GHz luminosity function of
  \cite{Mauch2007} converted to $325$ MHz assuming spectral index of $\alpha = -0.7$. We estimate that the lower limit on the local star formation
  rate density of the Universe is $\rho_{SFR} = 0.013  \pm 0.002$ M$_\odot$
  yr$^{-1}$ Mpc$^{-3}$ (Fig.~\ref{SFLF1}).           

\item We determine RLFs of radio-loud AGN for two redshift
  slices. Fitting a double power mild evolution out to $z \sim 0.5$. Parametrizing the evolution as $\propto
  (1+z)^{k}$, we find a best fitting values $k = 0.92 \pm 0.95$
  assuming pure number density and $2.13
  \pm 1.96$ assuming the pure luminosity density (Fig.~\ref{AGNEvolve}). From fitting
  single power law functions to the RLFs we show that low-luminosity
  sources evolve less than high-luminosity sources.    

\item We produce RLFs for the HERG and LERG populations of AGN. LERGs
  are the dominant population at luminosities between $
10^{23} < L_{\rm{325-MHz}} < 10^{26}$. At higher luminosities the
space densities of the populations become comparable (Fig.~\ref{HERGLERG}). 

\item After dividing the AGN sample into steep and flat-spectrum
  sources, we show that the extrapolation of the steep-spectrum
  sources agrees with the shifted flat-spectrum, as expected from the
  orientation-based unification of radio-loud AGN (Fig.~ \ref{FLATSTEEP}).    

\end{enumerate}

In order to fully understand the link between AGN-driven feedback and
the shutting down or continuous quenching of star formation in
galaxies, a full understanding of the link between radio power,
accretion mode and environment as a function of redshift is
required.
In the near future radio data obtained from the LOw-Frequency ARray
\citep[LOFAR;][]{VanHaarlem2013}, the Australian Square Kilometre
Array Pathfinder \citep[ASKAP;][]{Johnston2007} and MeerKAT
\citep{Jonas2009}, along with the VLA, will provide much greater
sample sizes for AGN and star forming galaxies
allowing more stringent constraints on the evolution of the radio
population, out to greater redshifts when combined
with future spectroscopic surveys \citep[e.g.][]{Smith2015}.     
Eventually, the Square Kilometre Array will be used to study the evolution of AGN from the cosmic dawn \citep{Smol2015}.     
Further work using the GAMA data set will involve investigating the evolution of HERG
and LERG populations of AGN at $1.4$ GHz (Pracy et al. (submitted)).    

\section {Acknowledgements}

MP, TM, KM, MJ, RJ and SF acknowledge support by the South African Square Kilometre Array
Project and the South African National Research Foundation.

GAMA is a joint European-Australasian project based
around a spectroscopic campaign using the Anglo-
Australian Telescope. The GAMA input catalogue is based
on data taken from the Sloan Digital Sky Survey and the
UKIRT Infrared Deep Sky Survey. Complementary imaging
of the GAMA regions is being obtained by a number of in-
dependent survey programs including GALEX MIS, VST KIDS, VISTA VIKING, WISE, Herschel-ATLAS, GMRT
and ASKAP providing UV to radio coverage. GAMA is
funded by the STFC (UK), the ARC (Australia), the AAO,
and the participating institutions. The GAMA website is http://www.gama-survey.org/

We acknowledge the IDL Astronomy User's Library, and IDL code maintained
by D.~Schlegel (IDLUTILS) as valuable resources.

We also thank the anonymous referee for their helpful comments which has improved the paper.  

\bibliographystyle{mnras}
\bibliography{Prescott15}

\bsp
\label{lastpage}

\end{document}